\documentclass[10pt,journal,comsoc]{IEEEtran}

\usepackage[normalem]{ulem}
\usepackage{multirow}
\usepackage{bbold}
\usepackage{ifpdf}
\usepackage{ulem} 
\usepackage{caption}
\usepackage{subcaption}%
\usepackage[cmex10]{amsmath}
\usepackage{amssymb}
\usepackage{verbatim}
\usepackage{algorithm}
\usepackage{algorithmic}
\usepackage{array}
\usepackage{latexsym}
\usepackage{color}
\usepackage{textgreek}
\usepackage{subscript}
\usepackage{rotating}
\usepackage{booktabs,multirow}
\usepackage{mdframed}	
\usepackage{tabularx}
\usepackage{amsmath}
\usepackage{makecell}
\usepackage{tabto}
\usepackage{mathrsfs}
\usepackage{url}
\usepackage{cite}
\usepackage{listings}
\usepackage{array}
\usepackage[table]{xcolor}
\usepackage{ragged2e}
\usepackage{booktabs}

\usepackage{graphicx}
\usepackage{comment}
\usepackage[cmex10]{amsmath}
\usepackage{amssymb}
\usepackage{bbold}
\usepackage{float}
\usepackage[cmex10]{amsmath}
\usepackage{amssymb}
\usepackage{verbatim}
\usepackage{array}
\usepackage{latexsym}
\usepackage{color}
\usepackage{tabularx} 
\usepackage{textgreek}
\usepackage{subscript}
\usepackage{rotating}
\usepackage{booktabs,multirow}
\usepackage{mdframed}	
\usepackage{tabularx}
\usepackage{amsmath}
\usepackage{makecell}
\usepackage{tabto}
\usepackage{mathrsfs}
\usepackage{url}
\usepackage{cite}
\usepackage{listings}
\usepackage{array}
\usepackage[table]{xcolor}

\definecolor{pinkpurple}{rgb}{0.6, 0.1, 0.9} 

\begin{document}

\title{Structure-Aware NL-to-SQL for SFC Provisioning via AST-Masking Empowered Language Models}
\author{
\IEEEauthorblockN{Xinyu Zhu, Parisa Fard Moshiri~\IEEEmembership{Student Member, IEEE}, Poonam Lohan~\IEEEmembership{SMIEEE},\\ Burak Kantarci~\IEEEmembership{SMIEEE},~and~Emil Janulewicz\vspace{-0.1in}}\\
\thanks{
X. Zhu, P. Fard Moshiri, P. Lohan and B. Kantarci are with the School of Electrical Engineering and Computer Science, University of Ottawa, Ottawa, ON, Canada. Emails: \{xzhu095, parisa.fard.moshiri,ppoonam,burak.kantarci\}@uottawa.ca

Emil Janulewicz is with Ciena, 383 Terry Fox Dr, Kanata, ON K2K 2P5, Canada, Email: ejanulew@ciena.com 
}
}


\IEEEtitleabstractindextext{
\begin{abstract}
Effective Service Function Chain (SFC) provisioning requires precise orchestration in dynamic and latency-sensitive networks. Reinforcement Learning (RL) improves adaptability but often ignores structured domain knowledge, which limits generalization and interpretability. Large Language Models (LLMs) address this gap by translating natural language (NL) specifications into executable Structured Query Language (SQL) commands for specification-driven SFC management. Conventional fine-tuning, however, can cause syntactic inconsistencies and produce inefficient queries. To overcome this, we introduce Abstract Syntax Tree (AST)-Masking, a structure-aware fine-tuning method that uses SQL ASTs to assign weights to key components and enforce syntax-aware learning without adding inference overhead. Experiments show that AST-Masking significantly improves SQL generation accuracy across multiple language models. FLAN-T5 reaches an Execution Accuracy (EA) of 99.6\%, while Gemma achieves the largest absolute gain from 7.5\% to 72.0\%. These results confirm the effectiveness of structure-aware fine-tuning in ensuring syntactically correct and efficient SQL generation for interpretable SFC orchestration.
\end{abstract}

}
\pagestyle{empty}

\maketitle
\thispagestyle{empty}
\IEEEdisplaynontitleabstractindextext
\IEEEpeerreviewmaketitle

\section{Introduction}

Modern networks increasingly rely on Software Defined Neworking (SDN) and Network Function Virtualization (NFV), which enable flexible resource management and service deployment\cite{WuShengSurvey}.  In this framework, Service Function Chain (SFC) provisioning links Virtual Network Functions (VNFs) into processing pipelines that support diverse, latency-sensitive services such as immersive media, cloud gaming, industrial automation, and large-scale IoT applications~\cite{WuShengSurvey}. The dynamic and resource-intensive nature of these services makes SFC provisioning complex and requires coordinated decisions on VNF placement, resource use, and execution order while meeting strict end-to-end performance goals.

Deep Learning (DL) has been extensively investigated for SFC provisioning, enabling the identification of intricate traffic patterns and enhancing orchestration efficiency~\cite{pei2020two}. It has been utilized for VNF resource forecasting, utilizing correlations among VNFs to enhance accuracy and facilitate resource allocation~\cite{kim2019deep}. RL has demonstrated potential in adaptive VNF placement and dynamic resource allocation amidst fluctuating traffic loads~\cite{chen2022}. Nevertheless, Reinforcement Learning (RL) solutions typically depend on structured state-action spaces and numerical inputs, which constrains their resilience in the face of unforeseen disruptions, such as outages or abrupt connection failures~\cite{parisa2025}. To overcome these limits, Deep RL (DRL) combines the feature extraction power of DL with the decision-making of RL. DRL has been applied to SFC orchestration and embedding, showing benefits for latency, energy efficiency, and online service provisioning in NFV-Edge Computing enabled IoT systems~\cite{pDRLproof}. However, DRL mainly works with numeric inputs and is less effective when dealing with unstructured signals or explaining its decisions.

 Language Models (LMs) offer a complementary approach by processing natural language queries and linking high-level service requirements to specific actions~\cite{nam2025}. Recent studies have integrated LMs with multi-objective optimization to support SFC deployment, jointly addressing latency, resource utilization, and energy efficiency while preserving semantic alignment~\cite{li2025}. However, applying LMs to operational querying, such as generating SQL to retrieve network states, faces a key challenge where unrestricted generation can introduce syntactic errors or semantic drift, undermining reliability and slowing the decision process. To address this, we propose an AST-Masking-assisted fine-tuning approach, which leverages SQL Abstract Syntax Trees (ASTs) to guide learning toward syntactically valid and semantically relevant tokens, thereby reducing errors and improving execution robustness. Building on our earlier Light Language Model (LiLM)-based Relational Data Base Assisted SFC Provisioning (LiLM-RDB-SFC) framework~\cite{parisa2025}, we integrate the AST-masked LiLM into a DRL-based SFC provisioning architecture, thereby enhancing responsiveness, resilience, and interpretability, and enabling precise and executable network state queries.

The main contributions of this paper are as follows:
\begin{enumerate}
    \item We propose an AST-Masking-assisted fine-tuning framework for LLMs with Low-Rank Adaptation (LoRA), enabling syntax-aware NL-to-SQL query generation tailored for network state monitoring and SFC provisioning.
    \item We develop a domain-specific NL-to-SQL dataset aligned with SFC relational schemas and evaluate Qwen, FLAN-T5, and Gemma under a unified setup.
    \item We demonstrate that heavy models are not essential for superior NL-to-SQL translation. Utilizing AST-Masking, lightweight language models attain accuracy comparable to that of larger models, enhancing deployability and inference efficiency in real-time SFC orchestration.
   
\end{enumerate}
AST-Masking yields consistent gains as follows: Exact Match (EM)/Execution Accuracy (EA)  reach 99.6\% for FLAN-T5 and 97.5\% for Qwen, while Gemma improves EM/EA from 7.5\% to 72.0\%. Furthermore, Valid Efficiency Score (VES), which is a metric indicating time, computation and database planner costs, improves by up to 13.2\% for Qwen, confirming syntactically valid and efficient SQL generation.

The remainder of this paper is organized as follows. Section II reviews related work, Section III presents the proposed methodology, Section IV provides performance evaluation and discussion, and Section V concludes the paper.

\section{Related Work}

 DL techniques have been extensively utilized in NFV and SDN contexts to tackle the issues of SFC provisioning and VNF placement.  Pei~\textit{et al.}~\cite{pei2020two} structured the VNF selection and chaining problem (VNF-SCP) as a Binary Integer Programming (BIP) model and introduced a DL-based Two-Phase Algorithm (DL-TPA) utilizing Deep Belief Networks (DBNs). Emu~\textit{et al.}~\cite{emu2020ensemble} proposed an ensemble DL framework for VNF deployment in IoT services, integrating Convolutional Neural Networks (CNNs) and Artificial Neural Networks (ANNs) trained with ILP-derived solutions. Nonetheless, DL-based approaches generally depend on offline training with structured datasets, which constrains their adaptability to real-time network variations.

RL has become a versatile substitute for adaptive NFV management.  Suzuki~\textit{et al.}~\cite{suzuki2020extendable} introduced an extensible NFV control framework that orchestrates many sub-controllers using a RL-based engine, enabling optimization across diverse metrics without the need for re-engineering the optimization pipeline.  Although RL enhances adaptability compared to static DL models, it frequently experiences extended exploration periods and diminished efficiency in managing intricate, multiple constraint DFC provisioning tasks.
 To improve flexibility, Chen~\textit{et al.}~\cite{chen2021drlqor} introduced DRL-QOR, a QoS/QoE-aware orchestration system designed as a Parameterized Action Markov Decision Process (PAMDP). Although DRL-QOR attains rapid convergence and enhanced long-term value, it predominantly focuses on numerical metrics and lacks semantic interpretability. 

 To mitigate these limitations, our previous research has explored the incorporation of LMs into DRL-based SFC provisioning.  In~\cite{parisa2025}, we presented the integration of numerical network states with semantic attributes derived from textual monitoring data, facilitating enhanced adaptability of DRL agents in dynamic situations.  In~\cite{parisa2025lilm}, we introduced LiLM-RDB-SFC, a framework that integrates a lightweight LM with a relational database to facilitate deep RL-based provisioning.
 
 This work presents a new expansion to the existing architecture: AST-masked fine-tuning for LMs.  AST masking imposes structural limitations during SQL creation, markedly enhancing syntactic validity while maintaining semantic freedom.  In comparison to previous researches, our AST-guided LM-DRL framework demonstrates enhanced SQL execution accuracy, generalization to unfamiliar query structures, and adaptability to changing network conditions, establishing it as a resilient solution for real-time NFV orchestration.
\begin{figure*}[!hbt]
       \centering
       \includegraphics[width = 0.9\textwidth, trim=0cm 0cm 0cm 0cm,clip]{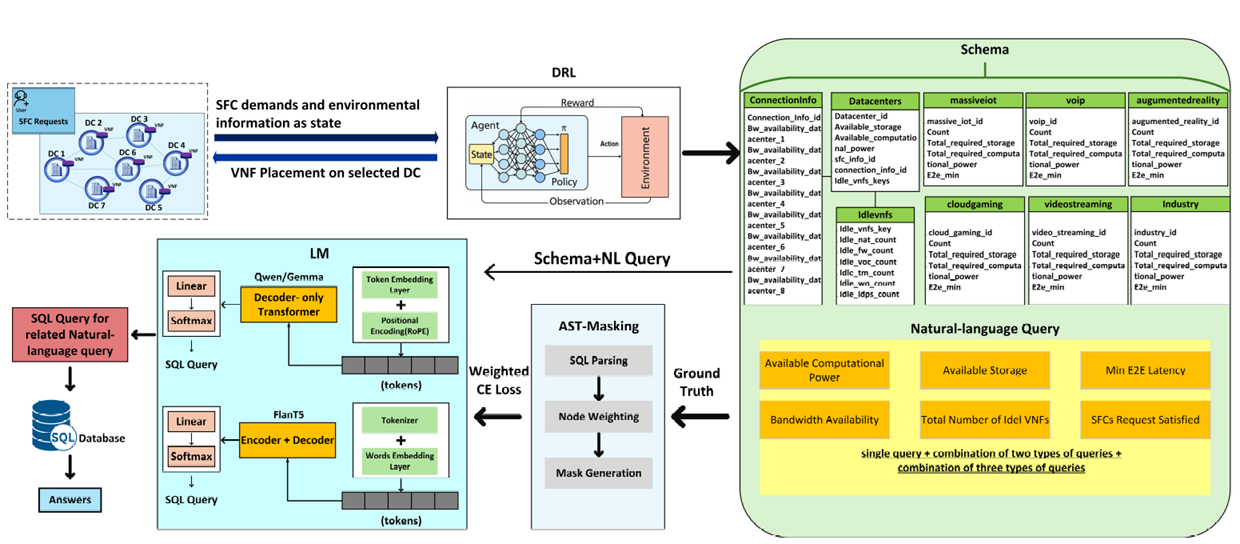}
       \caption{Framework Design for AST-Masking mechanism for fine-tuning LMs}
       \label{fig:system} 
\end{figure*}
\section{Methodology}
 In our previous study~\cite{arda24}, DRL was employed to optimize the acceptance rate of SFC requests within stringent infrastructure limitations.  The DRL agent successfully acquired optimal placement strategies customized for the varying resource and latency requirements of different SFCs.  While DRL can operate across different network configurations without retraining, it remains constrained when facing unpredictable network dynamics or incomplete state information.

To address these challenges, we establish a system that combines a language model with a relational SQL database for the dynamic administration of SFCs, DCs, and VNFs. This work focuses on the following VNFs: Network Address Translation (NAT), Intrusion Detection and Prevention System (IDPS), Video Optimization Controller (VOC), Firewall (FW), Traffic Monitor (TM), and WAN Optimizer (WO).   This study emphasizes the characterization and querying of fine-grained network metrics.  We specifically monitor critical metrics such end-to-end (E2E) latency for each SFC, idle VNF counts, resource utilization across DCs, and VNF processing delays.  The SQL database continuously updates these metrics following each DRL-driven activity, utilizing its robust design for efficient and reliable retrieval.  The LM operates as a NL interface, converting high-level NL quiries into SQL commands to retrieve metrics and facilitate network optimization decisions.

 This LM-SQL pipeline effectively bridges NL queries and database operations; however, the reliability of NL-to-SQL translation remains critical to maintaining system stability.  Structural inaccuracies in generated SQL, such as erroneous JOIN clauses, unbalanced parentheses, or discrepancies between tables and columns, can result in query failures that interrupt the feedback loop between the LM and the DRL agent.  To mitigate this vulnerability, we present \textbf{AST-Masking}, a structure-aware training enhancement that enhances the syntactic accuracy and validity of generated SQL queries. In this framework, the language model serves as a NL translation and monitoring interface rather than a policy-learning component of the DRL agent, and does not alter the agent’s reward or state transition dynamics.

 As illustrated in \figurename\hspace{0.1pt}\ref{fig:system}, comprehensive network state information is gathered after each action executed by the DRL agent. This state encompasses the existing SFC requirements, DC resources, link bandwidth, and latency information. The data collected is subsequently organized into a dynamic relational SQL database schema. NL queries, generated by users or automated monitoring systems, seek to obtain key performance indicators including: (i) available computational power, (ii) available storage, (iii) minimum E2E latency for a specific SFC at a DC, (iv) total number of idle VNFs, (v) bandwidth availability, and (vi) the number of satisfied SFC requests.
 
 The framework supports not only single-metric queries but also combined queries that involve two or three metrics, such as \{available storage and minimum E2E latency\}, or \{minimum E2E latency, available computational power and the total number of idle VNFs\}. The LM processes these NL queries with the related schema segments, converting them into executable SQL statements. Executing these SQL queries on the relational database obtains real-time network metrics, which are relayed to the monitoring interface. The AST-Masking mechanism, which improves LM performance through structure-aware training, is described next.

\subsection{AST-Masking}
\begin{figure*}[!hbt]
       \centering
       \includegraphics[width = 0.9\textwidth, trim=0cm 0cm 0cm 0cm,clip]{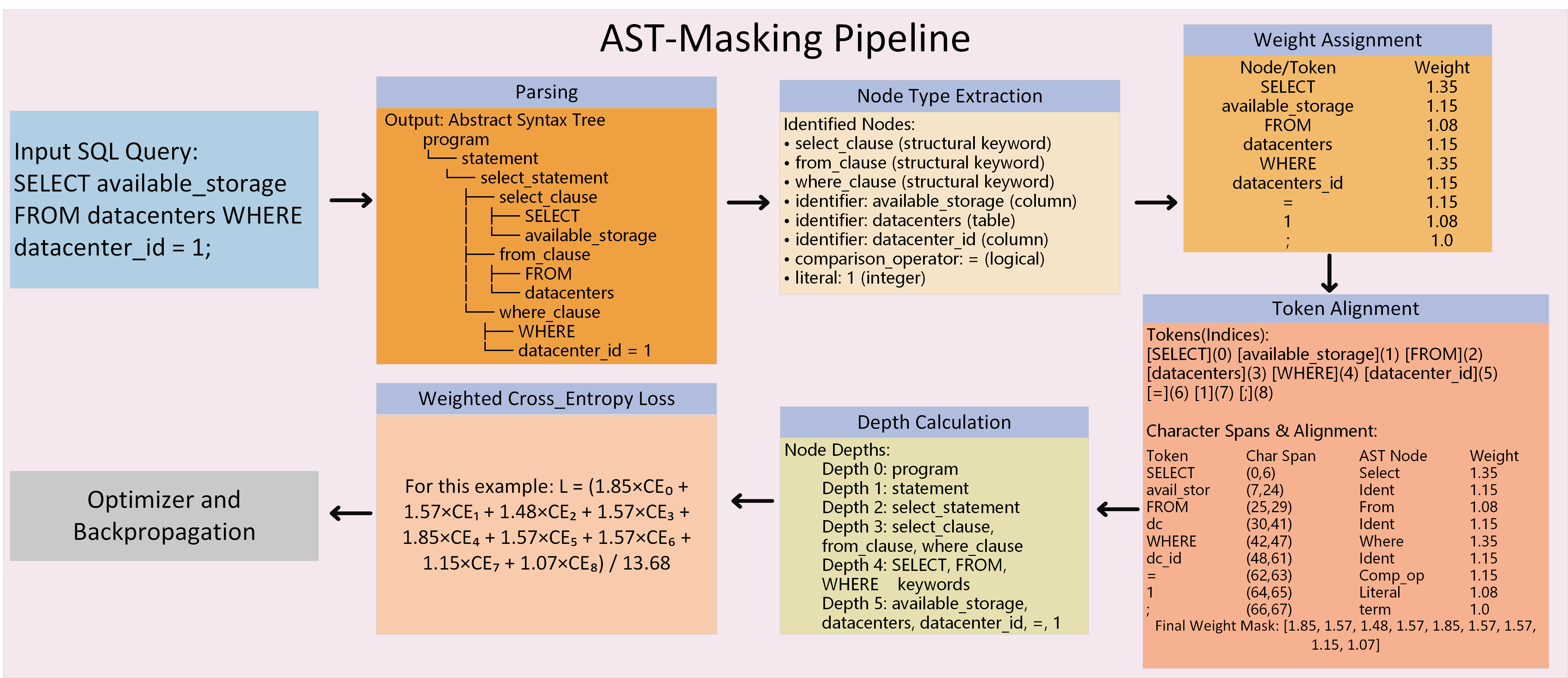}
       \caption{AST-Masking Pipeline}
       \label{fig:astmasking} 
\end{figure*}

To enhance the syntactic accuracy and execution success rate of SQL queries produced by the LM, we incorporate the AST-Masking module during the training phase.  As shown in Fig.~\ref{fig:astmasking}, AST-Masking parses each ground-truth SQL queries into its AST and assigns structure-dependent weights to individual tokens. These weights are then integrated into the cross-entropy(CE))loss, biasing the learning process toward syntactically critical elements (e.g., \texttt{SELECT}, \texttt{JOIN}, \texttt{WHERE}) while maintaining inference efficiency.
The overall process consists of six sequential steps:

\textbf{(1) SQL Parsing:} Each ground-truth SQL statement is parsed using the \textit{tree-sitter} library to construct an AST. The parser supports our customized SQL dialect, which includes nested subqueries, aggregation functions, and multi-table joins across the nine-table SFC schema (\texttt{ConnectionInfo}, \texttt{DataCenters}, \texttt{IdleVNFs}, and six SFC-specific tables).

\textbf{(2) Node Type Identification:} From the parsed AST, we identify the structural components most essential for syntactic correctness and semantic coherence. Empirical analysis of common generation errors reveals three major node categories:
(i) \textit{Critical structures}: query-defining clauses such as \texttt{SELECT} and \texttt{WHERE};
(ii) \textit{Important operations}: relational constructs such as \texttt{JOIN} and \texttt{GROUP BY};
(iii) \textit{Schema elements}: table and column identifiers corresponding to the network database.
    
\begin{figure*}[t]
  \centering
  \scalebox{0.9}{
    \begin{minipage}{\textwidth}
      \centering
      \begin{subfigure}[t]{0.32\textwidth}
        \centering
        \includegraphics[width=\linewidth]{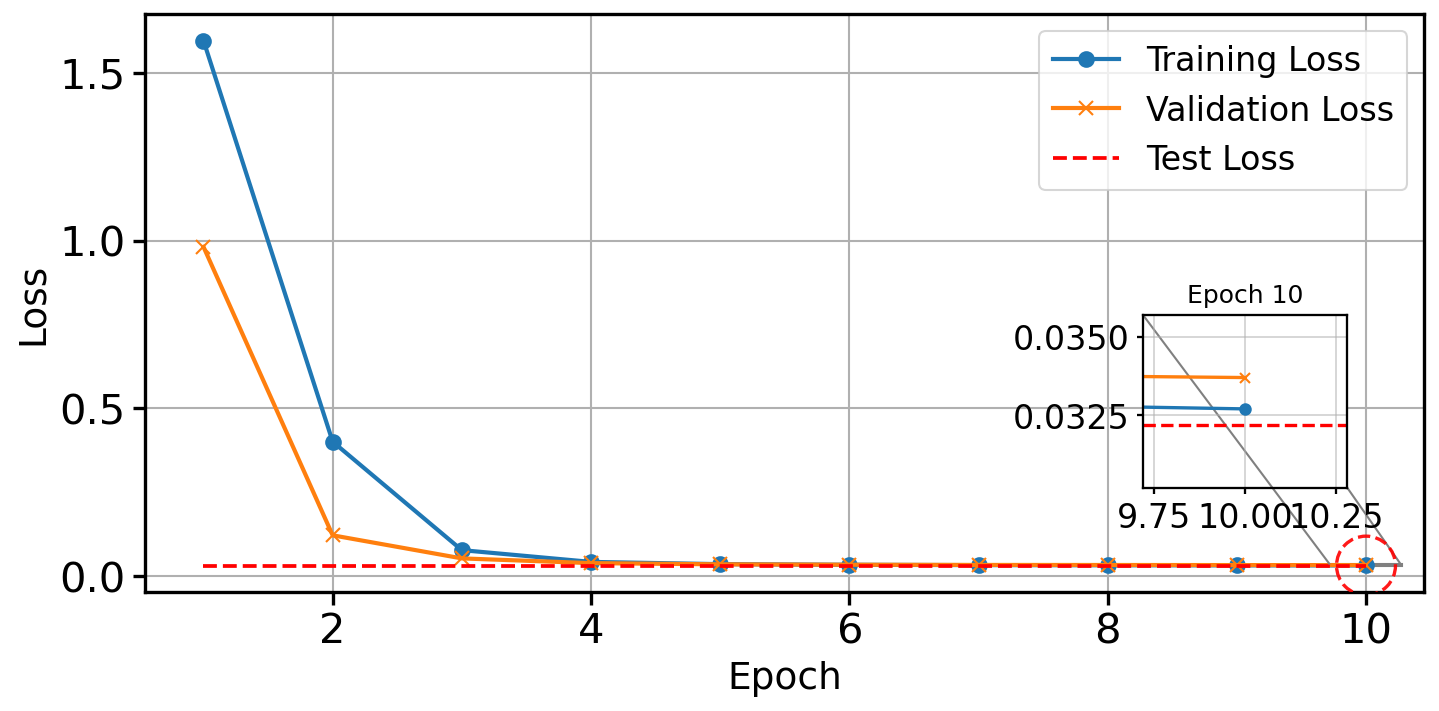}
        \caption{Qwen Baseline}
        \label{fig:qwen_base}
      \end{subfigure}\hfill
      \begin{subfigure}[t]{0.32\textwidth}
        \centering
        \includegraphics[width=\linewidth]{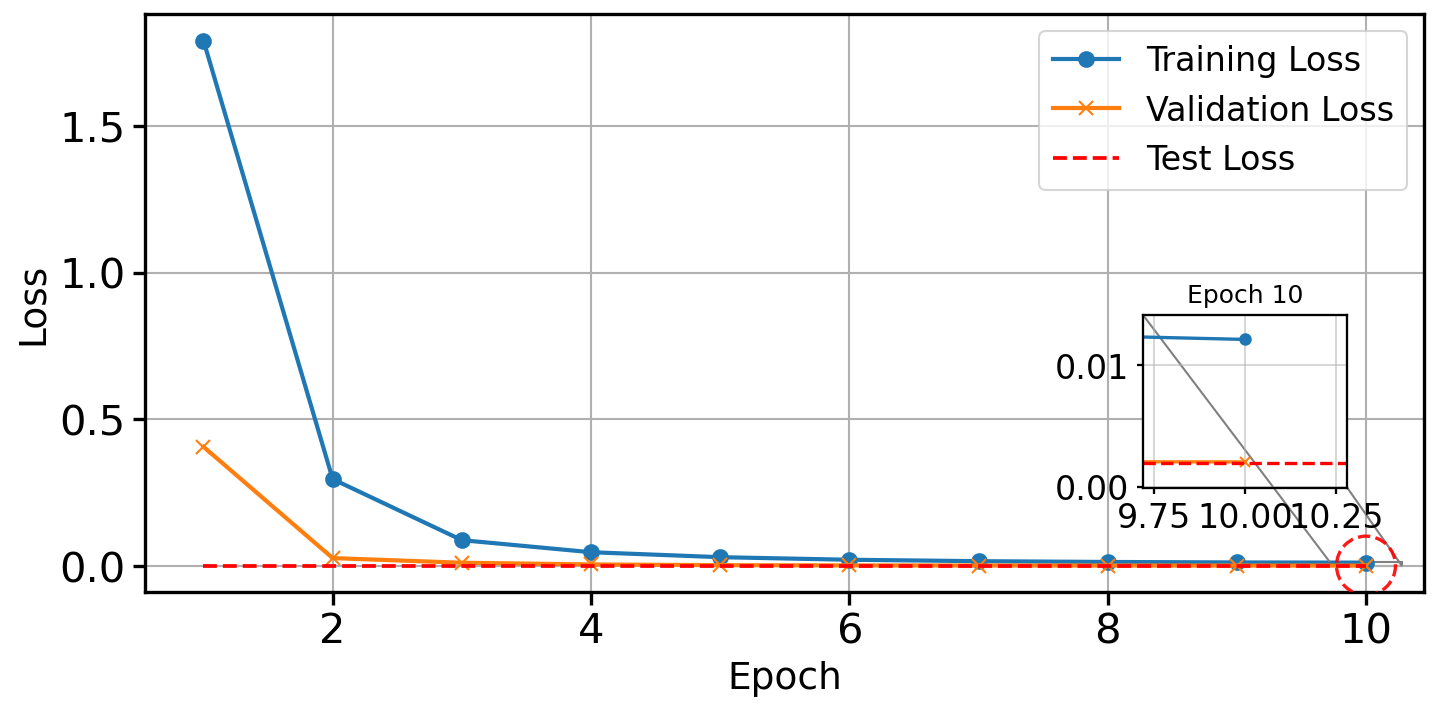}
        \caption{FLAN-T5 Baseline}
        \label{fig:flan_base}
      \end{subfigure}\hfill
      \begin{subfigure}[t]{0.32\textwidth}
        \centering
        \includegraphics[width=\linewidth]{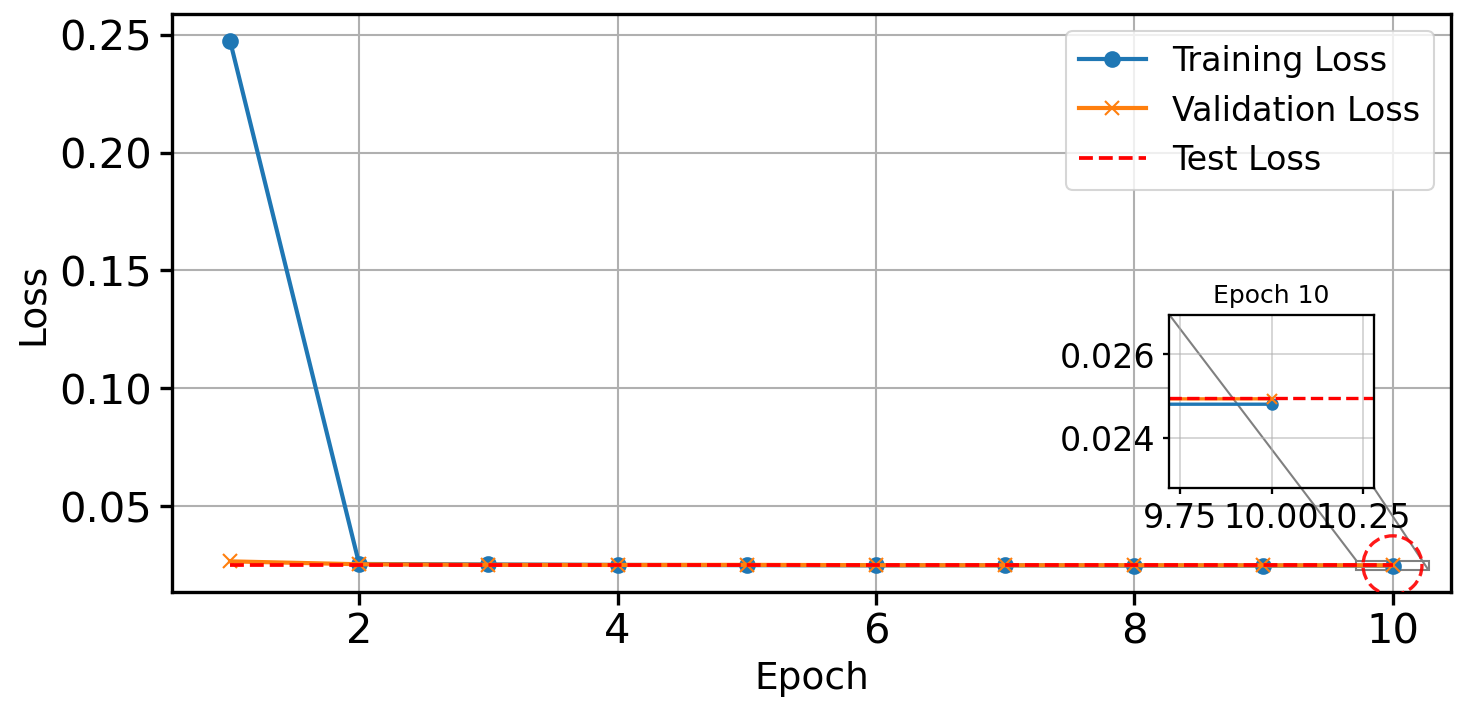}
        \caption{Gemma Baseline}
        \label{fig:gemma_base}
      \end{subfigure}

      \medskip

      \begin{subfigure}[t]{0.34\textwidth}
        \centering
        \includegraphics[width=\linewidth]{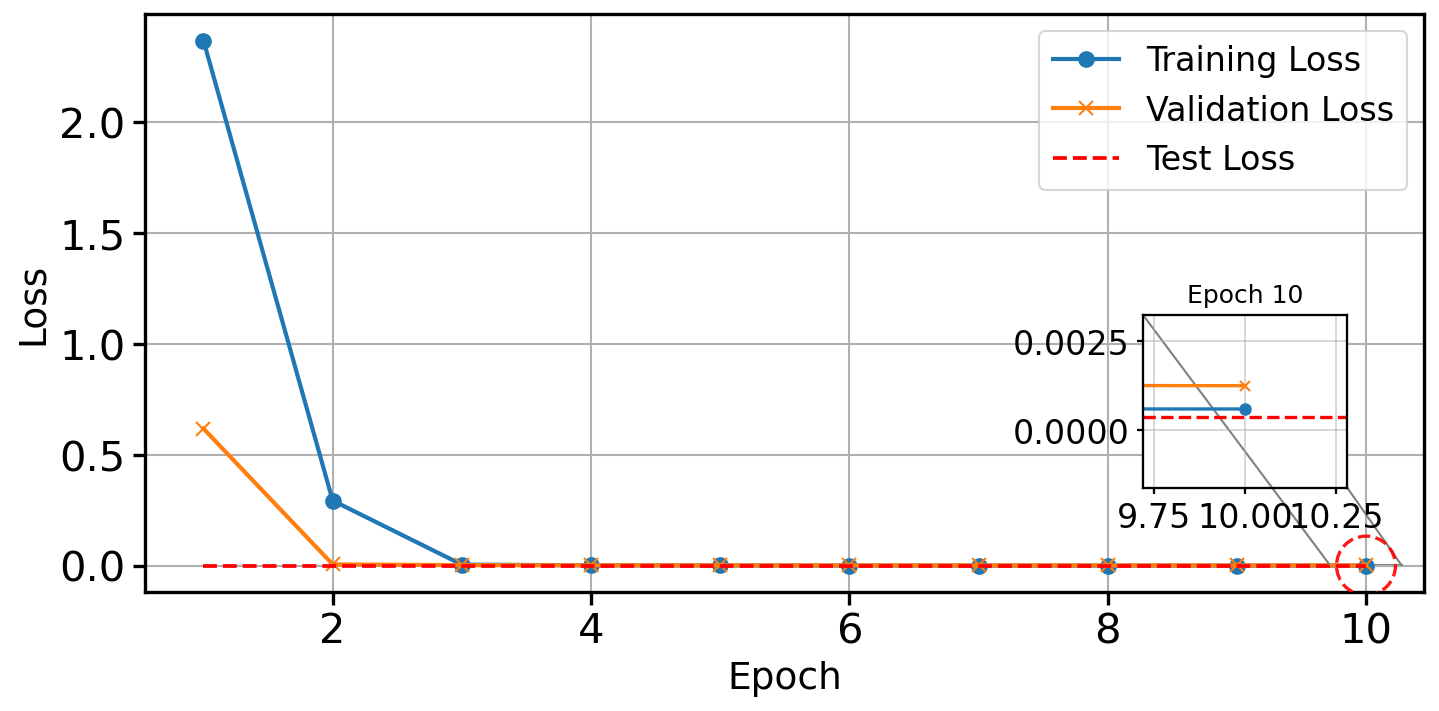}
        \caption{Qwen AST-Masking}
        \label{fig:qwen_ast}
      \end{subfigure}\hfill
      \begin{subfigure}[t]{0.32\textwidth}
        \centering
        \includegraphics[width=\linewidth]{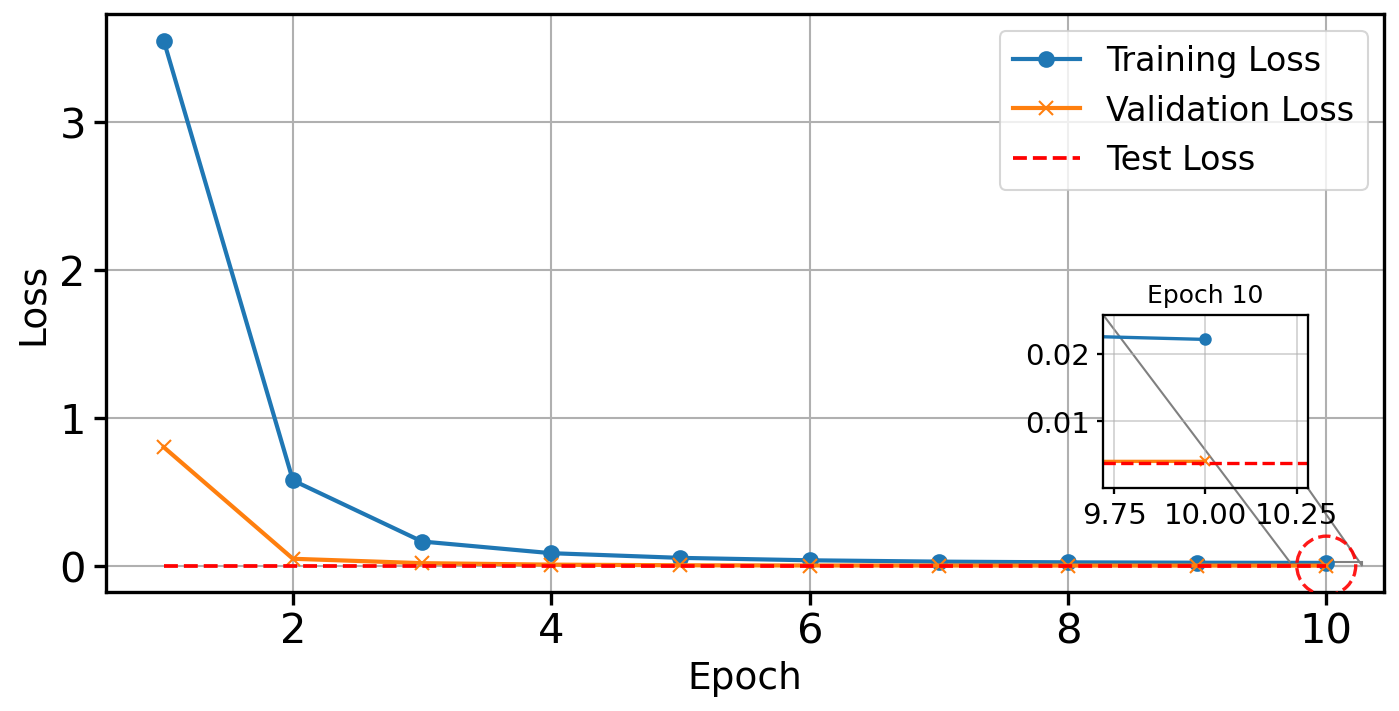}
        \caption{FLAN-T5 AST-Masking}
        \label{fig:flan_ast}
      \end{subfigure}\hfill
      \begin{subfigure}[t]{0.32\textwidth}
        \centering
        \includegraphics[width=\linewidth]{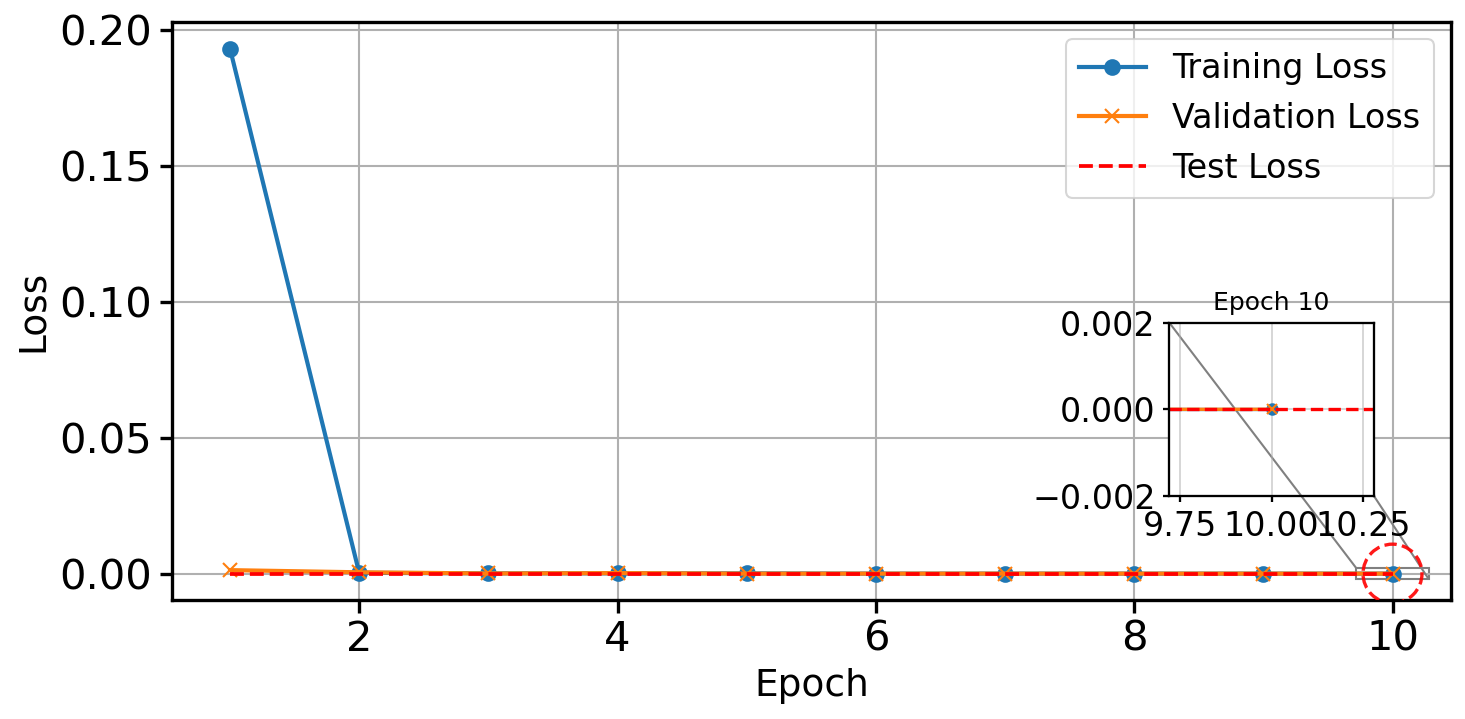}
        \caption{Gemma AST-Masking}
        \label{fig:gemma_ast}
      \end{subfigure}
    \end{minipage}
  }
  \caption{Loss curves of baseline vs.\ AST-Masking across model families.}
  \label{fig:loss_grid}
\end{figure*}

\textbf{(3) Weight Assignment:} 
 Each node type is assigned a significance coefficient, forming the set of \textit{AST-Masking weights}. Higher weights are applied to core clauses (\texttt{SELECT}, \texttt{FROM}, \texttt{WHERE}) and aggregate operators, while literals and constants receive smaller weights (see Fig.~\ref{fig:astmasking}). This weighting strategy biases optimization toward preserving SQL structure and reducing SQL syntax and execution errors.

\textbf{(4) Token Alignment:} 4) Token Alignment: Different tokenizers may produce varying token sequences for the same SQL expression. To ensure consistency, a character-level alignment maps each token to its corresponding AST node, independent of the tokenizer used. When a keyword such as “SELECT” is split into multiple subtokens, each inherits the original token’s structural weight (e.g., 1.35), ensuring uniform weighting and comparable behavior across model architectures.

\textbf{(5) Depth-Aware Adjustment:} We implement a hierarchical structure by normalizing the depth of nodes:

\noindent\textbf{Depth normalization:}
To incorporate hierarchical context, node `depth' is normalized as:
\begin{equation}
\label{eq:depth_norm}
d_{\text{norm}} \;=\; \min\!\left({\text{depth}}/{12},\, 1.0\right)
\end{equation}

\noindent\textbf{Token weighting:}
The token weight is computed as:
\begin{equation}
\label{eq:mask}
m_t \;=\; w_{\text{type}} \,\times\, \Bigl( 1 \,+\, 0.3\,s_{\text{mask}} \,+\, 0.2\,d_{\text{norm}} \Bigr)
\end{equation}
where $w_{\text{type}}$ is the base importance weight of a token, reflecting its structural role in the SQL query. It is computed by taking the higher value between the token’s AST node-type weight (e.g., for clauses like \texttt{SELECT} or \texttt{WHERE}) and its regex-based identifier weight (e.g., for table or column names).$s_{\text{mask}} \in \{0,1\}$ indicates whether the token, $t$, lies within a structural node (e.g., \texttt{select\_clause}, \texttt{where\_clause}, \texttt{join\_clause},\texttt{identifier}, etc.).

The token weights of each sample are normalized by mean over valid targets $V=\{t\mid \text{label}_t\neq -100\}$ to preserve the effective learning rate:
\begin{equation}
\label{eq:mean_norm}
\tilde{m}_t \;=\; \frac{m_t}{\frac{1}{|V|}\sum_{u\in V} m_u}, \qquad t\in V
\end{equation}

\textbf{(6)Training with Weighted  Loss:}
During training, we substitute the conventional CE with the AST-weighted alternative:
\begin{equation}\small
\label{eq:ast_loss_weighted}
\mathcal{L}_{\mathrm{AST}}
\;=\;
\frac{\sum_{t \in {V}} \tilde{m}_t \,\mathrm{CE}(y_t,\hat{y}_t)}
     {\sum_{t \in {V}} \tilde{m}_t}\,,
\qquad
V=\{\,t \mid \mathrm{label}_t\neq -100\,\},
\end{equation}

where $\tilde{m}_t$ is the normalized AST-derived weight for token $t$,
$y_t$ denotes the ground-truth token, $\hat{y}_t$ the model-predicted distribution,
and $\mathrm{CE}(\cdot,\cdot)$ is token-level cross-entropy. This formulation increases the penalty for structural predictions, steering the LM toward syntactically valid and executable SQL generation.

\subsection{ Language Models (LMs) under study}
\tablename\hspace{0.1pt} \ref{tab:model_rationale} summarizes the LMs considered in this study, outlining their architectural configurations and trade-offs between computational efficiency and representational capacity for NL-to-SQL generation.

 \textbf{Qwen} is a decoder-only transformer model created by Alibaba Cloud~\cite{qwen2025qwen25technicalreport}, intended for superior NL comprehension and generation across many tasks. Qwen, as a causal LM, demonstrates proficiency in autoregressive sequence creation, rendering it appropriate for text-to-SQL jobs that require syntactic and semantic coherence.

 \textbf{FLAN-T5} is an encoder-decoder transformer developed by Google Research~\cite{chung2022scalinginstructionfinetunedlanguagemodels}. It enhances the original T5 design by framing all activities as text-to-text problems and is instruction-tuned to increase generalization to unfamiliar tasks. The encoder acquires bidirectional contextual representations from the input sequence (NL query + schema), while the decoder produces the SQL output in an autoregressive fashion. In our previous research~\cite{parisa2025lilm}, FLAN-T5 demonstrated the highest Execution Accuracy and the lowest validation loss among the assessed lightweight models, hence prompting its incorporation in this study.
 

\textbf{Gemma} is a decoder-only transformer family introduced by Google~\cite{gemma_2025}, characterized by compact, open-weight models that are optimized for efficient fine-tuning and low-latency deployment.  As a causal language model, Gemma conditions on the concatenated NL query and schema, generating SQL tokens autoregressively, which is well-suited for text-to-SQL creation.
 All three models are trained using the proposed AST-Masking weighted cross-entropy (CE) loss.

\section{Performance Evaluation}

\begin{table}[t]
  \caption{Models considered and inclusion/exclusion rationale. }
  \label{tab:model_rationale}
  \centering
  \scriptsize
  \setlength{\tabcolsep}{3pt}
  \begin{tabularx}{\columnwidth}{l c >{\RaggedRight\arraybackslash}X}
    \toprule
    \textbf{Model / Params} & \textbf{Structure / Layers} & \textbf{Comments} \\
    \midrule
    Qwen2 / 500M & Decoder-only / 24 & \textit{Included}: Less hardware-constrained. \\
    FLAN-T5-Base / 248M & Encoder-decoder / 12-12 & \textit{Included}: Most effective in \cite{parisa2025lilm}. \\
    Gemma / 270M & Decoder-only / 18 & \textit{Included}: Alternative framework to Qwen. \\
    \addlinespace[1pt]
    BART-Base / 139M & Encoder-decoder / 6-6 & \textit{Excluded}: Weakest model in \cite{parisa2025lilm}. \\
    SQL-Coder / 7B & Decoder-only / 32 & \textit{Excluded}: High computational cost \cite{parisa2025lilm}. \\
    \bottomrule
  \end{tabularx}
  \vspace{-2mm}
\end{table}

The experiments are carried out on a system equipped with NVIDIA A100-PCIE-40GB GPUs, each featuring 40 GB of memory. The dataset comprises 38,536 manually crafted NL-SQL pairs with corresponding ground truth results and schema information. The dataset covers 36 query types combining six key metrics: available computational power, storage, idle VNFs, minimum E2E latency, bandwidth, and resource comparison. The data are split into 80\% for training, 10\% for validation, and 10\% for testing. The main fine-tuning hyperparameters for Qwen, FLAN-T5, and Gemma are summarized in \tablename\hspace{0.1pt}~\ref{tab:training_params}. The AST-Masking weights are empirically determined through extensive validation and ablation-based tuning, with the objective of enforcing relative structural importance rather than performing fine-grained numerical optimization.

Figs.~\ref{fig:loss_grid}\subref{fig:qwen_base}--\ref{fig:loss_grid}\subref{fig:gemma_ast} illustrate the training and validation loss trends for baseline and AST-Masked models under the CE loss. The introduction of AST-Masking enhances convergence stability across all model families by emphasizing structurally critical SQL components during optimization. In Qwen, validation loss decreases from 0.033 to 0.0013 and test loss from 0.032 to 0.0004, demonstrating faster and more stable convergence. FLAN-T5 exhibits a slight increase in loss after AST-Masking; however, this does not indicate degradation. The weighted loss re-prioritizes structural tokens (e.g., SELECT, JOIN, WHERE) and thus produces numerically higher but semantically more meaningful values. In contrast, Gemma shows the largest absolute gain, with validation and test losses dropping from 0.0249 to $6.76\times 10^{-8}$ and $9.12\times 10^{-8}$, respectively. Minor cases where training loss exceeds validation or test loss arise from differences between mini-batch training with dropout and averaged evaluation without dropout.
\begin{table}[t]
\centering
\caption{Final hyperparameter configurations used for fine-tuning baseline (B) and AST-Masking (A) models}
\resizebox{\columnwidth}{!}{ 
\begin{tabular}{lcccccc}
\toprule
\textbf{Param.} & \textbf{Qwen-B} & \textbf{Qwen-A} & \textbf{FLAN-T5-B} & \textbf{FLAN-T5-A} & \textbf{Gemma-B} & \textbf{Gemma-A} \\
\midrule
Learning rate & $1\times 10^{-5}$ & $1\times 10^{-5}$ & $1\times 10^{-5}$ & $1\times 10^{-5}$ & $2\times 10^{-5}$ & $1\times 10^{-5}$ \\
Batch size & 8 & 8 & 4 & 4 & 8 & 8 \\
LoRA rank $r$ & 8 & 8 & 8 & 8 & N/A & N/A \\
LoRA $\alpha$ & 8 & 8 & 8 & 8 & N/A & N/A \\
LoRA dropout & 0.05 & 0.05 & 0.05 & 0.05 & N/A & N/A \\
\bottomrule
\textit{* B: Baseline, A: AST-Masking}
\label{tab:training_params}
\end{tabular}}
\end{table}

\begin{table}[t]
  \centering
  \caption{Performance comparison }
  \label{tab:performance}
  \scriptsize
  \setlength{\tabcolsep}{3pt}
  \renewcommand{\arraystretch}{1.1}
  \begin{tabularx}{\columnwidth}{l *{5}{>{\centering\arraybackslash}X}}
    \toprule
    \multirow{2}{*}{\textbf{Model}} & \textbf{EM} & \textbf{EA} & \textbf{AvgTime} & \textbf{AvgComplex} & \textbf{VES} \\
     & (\%) & (\%) & (\%) & (\%) & (\%) \\
    \midrule
    Qwen-B   & 83.9 & 83.9 & 91.7 & 100.0 & 81.2 \\
    Qwen-A   & 97.5 & 97.5 & 91.7 & 100.0 & 94.4 \\
    FLAN-T5-B & 94.1 & 94.1 & 89.9 & 100.0 & 90.3 \\
    FLAN-T5-A & 99.6 & 99.6 & 92.3 & 100.0 & \textbf{96.5} \\
    Gemma-B  & \textbf{7.5}  & \textbf{7.5} & 92.7 & 100.0 & 7.3 \\
    Gemma-A  & \textbf{72.0} & \textbf{72.0} & 91.9 & 100.0 & 69.7 \\
    \bottomrule
  \end{tabularx}
\end{table}

Further, We define a set of evaluation metrics to assess the performance of Qwen, FLAN-T5 and Gemma under both baseline and AST-Masking-assisted fine-tuning.
\subsubsection{Execution Accuracy (EA)}
Let $\mathrm{Exec}(\cdot)$ denote the result set returned by executing a SQL query on PostgreSQL. Let $N$ denotes the number of evaluated queries (test NL-SQL pairs), and $i$ indexes the $i$-th query.
We regard a prediction $\hat{y}_i$ as correct if it executes without error and matches the ground-truth $y_i$ in
(a) number of rows, (b) number of columns, and (c) multiset of values (order-insensitive).
\begin{equation}\small
\mathrm{EA}=\frac{1}{N}\sum_{i=1}^{N}\mathbf{1}\!\left[\mathrm{Exec}(\hat{y}_i)=\mathrm{Exec}(y_i)\right]
\end{equation}
\subsubsection{Average Time (Avg\_Time)}
For each query $i$, we measure the average fetch latency $\bar t^{(i)}$ (in sec) over $K$ runs ($K{=}5$)
after a warm up execution. We report the dataset average.

\subsubsection{Average Complexity (Avg\_Complex)}
We compute a heuristic SQL complexity score as a weighted sum over structural features extracted from the query:
\begin{equation}\small
C = \sum_{f\in\mathcal{F}} w_f \cdot \mathrm{count}_f
\end{equation}
where $\mathcal{F}$ is the set of structural features considered.
We calculate a per query SQL complexity score by enumerating selected structural features and aggregating these counts with predetermined integer weights. (e.g., \texttt{COUNT}/\texttt{SUM}/\texttt{AVG}/\texttt{MAX}/\texttt{MIN}), window functions, and \texttt{CASE WHEN} constructs; the default weights are $w_{\text{joins}}{=}3$, $w_{\text{subqueries}}{=}4$, $w_{\text{group\_by}}{=}2$, $w_{\text{order\_by}}{=}1$, $w_{\text{having}}{=}2$, $w_{\text{distinct}}{=}1$, $w_{\text{union}}{=}3$, $w_{\text{aggregations}}{=}1$, $w_{\text{window}}{=}3$, $w_{\text{case\_when}}{=}2$.
We empirically determined these integer weights through trial-and-error.

\subsubsection{Valid Efficiency Score (VES)}
For a correct prediction ($\mathrm{Exec}(\hat{y}_i){=}\mathrm{Exec}(y_i)$), we define three relative efficiency sub-scores against the ground truth (GT):
\begingroup
\small
\begin{align}
S^{(i)}_{\text{time}} &= 
\begin{cases}
1, & \bar t^{(i)}_{\mathrm{pred}} \le \bar t^{(i)}_{\mathrm{GT}} \\
\max\!\big(0.1, \ \frac{\bar t^{(i)}_{\mathrm{GT}}}{\bar t^{(i)}_{\mathrm{pred}}}\big), & \text{otherwise}
\end{cases} \\
S^{(i)}_{\text{comp}} &= 
\begin{cases}
1, & C^{(i)}_{\mathrm{pred}} \le C^{(i)}_{\mathrm{GT}} \\
\max\!\big(0.1, \ \frac{C^{(i)}_{\mathrm{GT}}}{C^{(i)}_{\mathrm{pred}}}\big), & \text{otherwise}
\end{cases} \\
S^{(i)}_{\text{plan}} &= 
\begin{cases}
1, & \mathrm{Cost}^{(i)}_{\mathrm{pred}} \le \mathrm{Cost}^{(i)}_{\mathrm{GT}} \\
\max\!\big(0.1, \ \frac{\mathrm{Cost}^{(i)}_{\mathrm{GT}}}{\mathrm{Cost}^{(i)}_{\mathrm{pred}}}\big), & \text{otherwise}
\end{cases}
\end{align}
\endgroup
 where \(S^{(i)}_{\text{time}}\) is time efficiency, evaluating the average execution duration of the expected query against the actual query, as queries may vary in their execution times on the database. 
\(S^{(i)}_{\text{comp}}\) represents complexity efficiency and denotes the comparative structural simplicity of the query, determined by weighted SQL constructs (e.g., JOIN = 3, SUBQUERY = 4). For instance, one query may use only a simple \texttt{SELECT} with \texttt{WHERE}, while another may rely on nested subqueries or multiple \texttt{JOIN}s. This metric is important because simpler queries are generally easier to maintain, optimize, and execute efficiently.
\(S^{(i)}_{\text{plan}}\), plan efficiency, evaluates the expenses associated with the database planner, such as PostgreSQL's \texttt{Total Cost}. In PostgreSQL, the planner generates an execution plan and attaches a numeric “Total Cost” that reflects the expected resources required (CPU, I/O, memory) for executing the query. 
Additionally, $\bar t$ is the average fetch latency, $C$ represents the aforementioned complexity score above, and $\mathrm{Cost}$ is PostgreSQL's \texttt{EXPLAIN (FORMAT JSON)} total Cost.
Considering nonnegative weights $w_{\text{time}}{+}w_{\text{comp}}{+}w_{\text{plan}}{=}1$ (in our case $w_{\text{time}}{=}0.4$, $w_{\text{comp}}{=}0.3$, $w_{\text{plan}}{=}0.3$), the per-query VES is
\begin{equation}\small
\mathrm{VES}^{(i)}=
\begin{cases}
w_{\text{time}}S^{(i)}_{\text{time}} + w_{\text{comp}}S^{(i)}_{\text{comp}} + w_{\text{plan}}S^{(i)}_{\text{plan}}, & \text{if correct} \\
0, & \text{otherwise}
\end{cases}
\end{equation}

\noindent Table~\ref{tab:performance} presents the comparative performance of baseline and AST-Masked models across five evaluation metrics: EM, EA, AvgTime, AvgComplex, and VES. EM and EA yield identical values because each NL-SQL pair in our schema corresponds to a single deterministic reference. Thus, syntactic correctness ensures identical execution results, making EM and EA coincide in this controlled evaluation. AST-Masking consistently improves EM and EA across Qwen and FLAN-T5, with Qwen increasing by +13.6\% and FLAN-T5 reaching 99.6\%, an improvement of +4.1\% over its already high baseline.  The AvgTime also remains nearly unchanged, demonstrating that AST-Masking improves accuracy without adding computational overhead. AvgComplex remains constant at 100\% for all models, confirming that structural weighting does not alter the underlying SQL design patterns. Notably, VES exhibits the largest relative gain, rising from 81.2\% to 94.4\% for Qwen and from 90.3\% to 96.5\% for FLAN-T5. Gemma shows the most substantial absolute improvement, with EM/EA increasing from 7.5\% to 72.0\% and VES from 7.3\% to 69.7\%. Overall, AST-Masking enhances both syntactic precision and execution efficiency across all model families.

\section{Conclusion}

This study has presented an AST-masked fine-tuning methodology that enhances the SQL generation proficiency of LMs for SFC provisioning.  The strategy improves syntactic validity and execution robustness by restricting learning using AST-derived token weights, without incurring inference-time overhead. Experiments reveal significant improvements: for Qwen, EM/EA increases from 83.9\% to 97.5\%, whilst for FLAN-T5, EM/EA enhance from 94.1\% to 99.6\%, approaching flawless accuracy.  VES demonstrates significant increases, rising from 81.2\% to 94.4\% in Qwen and from 90.3\% to 96.5\% in FLAN-T5, although query latency and complexity remain mostly unaltered. For Gemma, beginning from a lower baseline, the improvements are particularly notable, with EM/EA increasing from 7.5\% to 72.0\% and VES from 7.3\% to 69.7\%. These findings underscore the framework's capacity to concurrently enhance accuracy and efficiency across many model families. The AST-masked model functions as a semantic interface to the DRL agent within our comprehensive SFC orchestration pipeline, facilitating interpretable and query-driven state awareness for adaptive decision-making. 

Future research will utilize this methodology for more extensive network automation tasks. As the existing AST-Masking weights are fixed after post-empirical tuning, our research agenda includes the development of adaptive or learnable weighting algorithms to enhance portability across database schemes and SQL dialects.


\section*{Acknowledgment}
This work is supported by the Natural Sciences and Engineering
Research Council of Canada (NSERC) Alliance Program, MITACS
Accelerate Program under Project IT32016, and NSERC CREATE TRAVERSAL program.
\bibliographystyle{IEEEtran}


\end{document}